\documentclass{midl} 

\usepackage{mwe} 
\usepackage{amssymb}
\usepackage[skip=0pt]{caption}

\jmlrproceedings{MIDL}{Medical Imaging with Deep Learning}
\jmlrpages{}
\jmlryear{2023}

\jmlrworkshop{Short Paper -- MIDL 2023 submission}
\jmlrvolume{-- Under Review}
\editors{Under Review for MIDL 2023}

\title[CT Field-of-view Completion with Generative Diffusion Prior]{Zero-shot CT Field-of-view Completion with Unconditional Generative Diffusion Prior}

\midlauthor{\Name{Kaiwen Xu\nametag{$^{1}$}} \Email{kaiwen.xu@vanderbilt.edu}\\
  \Name{Aravind R. Krishnan\nametag{$^{1}$}} \Email{aravind.r.krishnan@Vanderbilt.edu}\\
  \Name{Thomas Z. Li\nametag{$^{1}$}} \Email{thomas.z.li@vanderbilt.edu}\\
  \Name{Yuankai Huo\nametag{$^{1}$}} \Email{yuankai.huo@vanderbilt.edu}\\
  \Name{Kim L. Sandler\nametag{$^{2}$}} \Email{kim.sandler@vumc.org}\\
  \Name{Fabien Maldonado\nametag{$^{2}$}} \Email{fabien.maldonado@vumc.org}\\
  \Name{Bennett A. Landman\nametag{$^{1,2}$}} \Email{bennett.landman@vanderbilt.edu}\\
  \addr {$^1$Vanderbilt University, 2301 Vanderbilt Place, Nashville, 37235, United States
    \\$^2$Vanderbilt University Medical Center, 1211 Medical Center Drive, Nashville, 37232, United States}}

\begin{document}

\maketitle

\begin{abstract}
  Anatomically consistent field-of-view (FOV) completion to recover
  truncated body sections has important applications in quantitative
  analyses of computed tomography (CT) with limited FOV. Existing
  solution based on conditional generative models relies on the
  fidelity of synthetic truncation patterns at training phase, which
  poses limitations for the generalizability of the method to potential
  unknown types of truncation. In this study, we evaluate a zero-shot
  method based on a pretrained unconditional generative diffusion
  prior, where truncation pattern with arbitrary forms can be
  specified at inference phase. In evaluation on simulated chest CT
  slices with synthetic FOV truncation, the method is capable of
  recovering anatomically consistent body sections and subcutaneous
  adipose tissue measurement error caused by FOV truncation. However,
  the correction accuracy is inferior to the conditionally trained
  counterpart.
\end{abstract}


\section{Introduction}

Quantitative analysis of medical images is less effective when body
sections of interest are partially truncated by limited imaging
field-of-view (FOV). This is especially an issue for opportunistic
assessment of body compositions using routine chest computed
tomography (CT) \citep{Troschel2020, Xu2021, Luo2021, Xu2022}.  A
previous study achieved anatomically consistent FOV extension of chest CT
by training a generative model conditioned on synthetic truncation
patterns \citep{xu2022body}. However, the effectiveness of this
approach heavily relies on the fidelity of simulated truncation
patterns, which makes it difficult to generalize to applications with
truncation patterns that are not considered in the simulation.  Recent
studies have demonstrated the possibility for zero-shot sampling of
semantic plausible images conditioned on partially corrupted image
data using an unconditionally trained generative diffusion prior
\citep{lugmayr2022repaint, Fei2023}. The conditioning information is
only needed at the inference, making it extremely flexible for
applications when corruption patterns are difficult to predict.

In this pilot study, we developed RePaint-DDIM, a variant of the
RePaint framework proposed by \cite{lugmayr2022repaint} modified for
the reverse sampling scheme of denoising diffusion implicit models
(DDIM) \citep{song2021denoising} and evaluated the method for the
task of FOV completion of routine lung screening low-dose CT.

\section{Method}

A generative diffusion model is trained to reverse a forward Markovian
diffusion process that progressively turns an image into noise. Once
developed, the model is capable of sampling a realistic image from
random noise following a recurrent procedure. Compared to the
original sampling scheme of denoising diffusion probabilistic model
(DDPM) \citep{ho2020denoising}, the DDIM reverse sampling scheme turns
the generative process into a deterministic process and reduces the
required sampling steps with a shortened sampling trajectory
\citep{song2021denoising}.

To condition the reverse sampling process on known image regions,
RePaint \citep{lugmayr2022repaint} alters the process by iteratively
replacing known regions of the intermediate reverse sampled image with
forward sampled image at each step of the DDPM denoising process. To
address the disharmony of the generated parts of the image, an additional
resampling process is introduced by iterating forward diffusion,
known region replacing, and denoising between
adjacent sampling steps.  \setlength{\textfloatsep}{0pt}
\begin{figure}[htbp]
  \floatconts
      {fig:method}
      {\caption{Overview of the mechanism for conditional field-of-view
          completion based-on unconditionally pre-trained generative diffusion
          prior.
      }}
      {\includegraphics[width=0.9\linewidth]{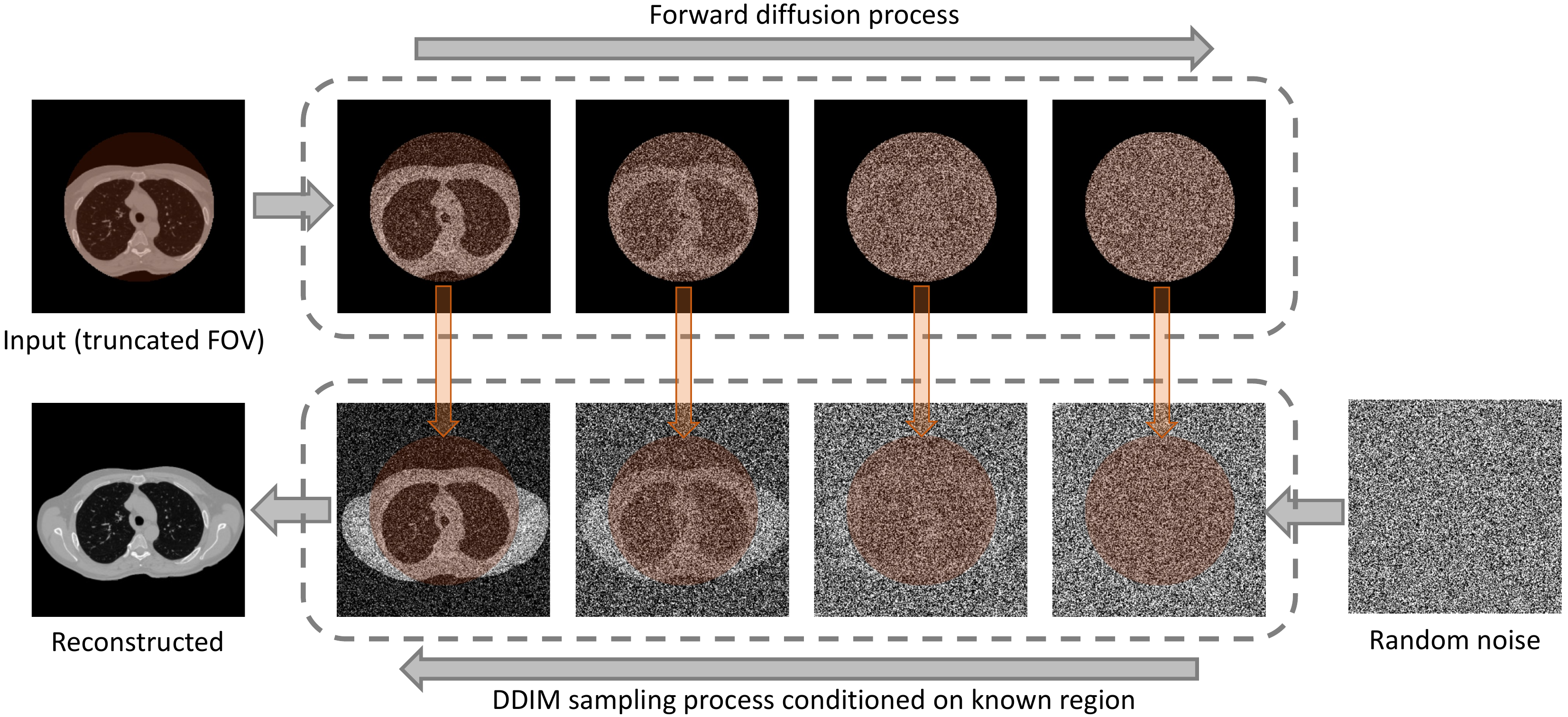}}
\end{figure}
In this study, we modified the RePaint algorithm for the DDIM sampling
process to take advantage of the shortened inference time. In the
resampling steps, instead of iterating between adjacent denoising
steps, the modified algorithm iterates between the predicted fully
denoised image ($x_0$) and intermediate sampled noisy image
($x_t$) in each DDIM sampling step.
We called this modified version as RePaint-DDIM. An overview of
the sampling workflow is demonstrated in Figure \ref{fig:method}.
Detailed steps are provided in Algorithm \ref{alg:inference}.
\setlength{\textfloatsep}{10pt}
\begin{algorithm2e}
  \small
  \caption{\small RePaint-DDIM sampling algorithm for CT field-of-view completion.}
  \label{alg:inference}
  \KwIn{$\tilde{x}_0$, CT slice with FOV truncation; $m$, FOV region mask; $\alpha_0:\alpha_T \in (0,1]$, diffusion schedule; $\epsilon_\theta$, pretrained denoising model.}
  \KwOut{$x_0$, CT slice with completed FOV}
  $x_T\leftarrow \mathcal{N}(0, I)$\;
  \For{$t\leftarrow T$ \KwTo $1$}{
    \For{$u\leftarrow 1$ \KwTo $U$}{
      $\hat{x}_0 \leftarrow \frac{1}{\sqrt{\alpha_t}} \left( x_t - \sqrt{1 - \alpha_t} \epsilon_\theta (x_t;t) \right)$\;
      $\hat{x}_0 \leftarrow m \odot \tilde{x}_0 + (1 - m) \odot \hat{x}_0$\;
      \If{$u < U$}{
        $\epsilon \leftarrow \mathcal{N}(0, I)$\;
        $x_t\leftarrow \sqrt{\alpha_t}\hat{x}_0 + \sqrt{1 - \alpha_t}\epsilon$\;
      }
    }
    $x_{t-1} \leftarrow \sqrt{\alpha_{t-1}}\hat{x}_0 + \sqrt{1-\alpha_{t-1}}\epsilon_{\theta}(x_t; t)$\;
}
\end{algorithm2e}
\section{Experiment and Discussion}
We pretrained an unconditional DDPM using 71,319 lung cancer screening
low-dose CT slices with complete body in FOV. Details of the
collection of this dataset were provided in \citep{xu2022body}. Slices
were resized to $256 \times 256$ and clipped to HU range $[-1000,
  600]$.  The model was trained with diffusion steps $T=1000$, linear
beta scheduler, and a batch size of 24. The model was trained for
30,000 iterations. At inference, we use 50 denoising steps and 20
resampling steps for each denoising step.

We evaluated the RePaint-DDIM on 2,657 simulated FOV truncation
slices generated from 145 withheld slices with complete body in
FOV. The anatomical consistency of the synthetic body sections was
quantitatively evaluated by the agreement of subcutaneous adipose
tissue (SAT) measured on reconstructed slices with the same
measurement on untruncated version following the method used in
\citep{xu2022body}. We compared the RePaint-DDIM with a
conditionally trained model as developed in \citep{xu2022body} (termed
S-EFOV). The results are provided in Figure \ref{fig:result}. The
method is capable of restoring anatomical consistent body sections in
the truncated region and correct the measurement error of
SAT. However, the correction accuracy is inferior to the conditionally
trained counterpart.
\begin{figure}[htbp]
  \floatconts {fig:result}
      {\caption{Evaluation of the anatomical consistency of the
          field-of-view (FOV) completion results. Tissue truncation index (TCI)
          reflects the severity of synthetic FOV truncation.}}
      {\includegraphics[width=.95\linewidth]{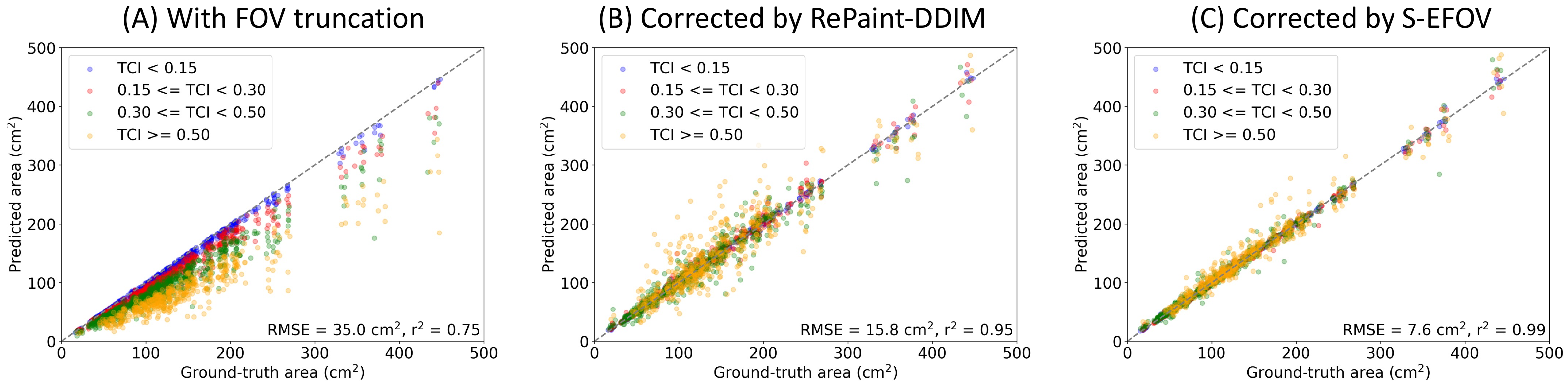}}
\end{figure}
\clearpage
\bibliography{ref}

\begin{thebibliography}{9}
\providecommand{\natexlab}[1]{#1}
\providecommand{\url}[1]{\texttt{#1}}
\expandafter\ifx\csname urlstyle\endcsname\relax
  \providecommand{\doi}[1]{doi: #1}\else
  \providecommand{\doi}{doi: \begingroup \urlstyle{rm}\Url}\fi

\bibitem[Fei et~al.(2023)Fei, Lyu, Pan, Zhang, Yang, Luo, Zhang, and
  Dai]{Fei2023}
Ben Fei, Zhaoyang Lyu, Liang Pan, Junzhe Zhang, Weidong Yang, Tianyue Luo,
  Bo~Zhang, and Bo~Dai.
\newblock Generative diffusion prior for unified image restoration and
  enhancement.
\newblock 4 2023.
\newblock URL \url{http://arxiv.org/abs/2304.01247}.

\bibitem[Ho et~al.(2020)Ho, Jain, and Abbeel]{ho2020denoising}
Jonathan Ho, Ajay Jain, and Pieter Abbeel.
\newblock Denoising diffusion probabilistic models.
\newblock \emph{Advances in Neural Information Processing Systems},
  33:\penalty0 6840--6851, 2020.

\bibitem[Lugmayr et~al.(2022)Lugmayr, Danelljan, Romero, Yu, Timofte, and
  Van~Gool]{lugmayr2022repaint}
Andreas Lugmayr, Martin Danelljan, Andres Romero, Fisher Yu, Radu Timofte, and
  Luc Van~Gool.
\newblock Repaint: Inpainting using denoising diffusion probabilistic models.
\newblock In \emph{Proceedings of the IEEE/CVF Conference on Computer Vision
  and Pattern Recognition}, pages 11461--11471, 2022.

\bibitem[Luo et~al.(2021)Luo, Terry, Tang, Xu, Massion, Landman, Carr, and
  Huo]{Luo2021}
Can Luo, James Terry, Yucheng Tang, Kaiwen Xu, Pierre Massion, Bennett~A.
  Landman, Jeffrey Carr, and Yuankai Huo.
\newblock Measure partial liver volumetric variations from paired
  inspiratory-expiratory chest ct scans.
\newblock page 112. SPIE, 2 2021.
\newblock ISBN 9781510640214.
\newblock \doi{10.1117/12.2581077}.
\newblock URL
  \url{https://www.spiedigitallibrary.org/conference-proceedings-of-spie/11596/2581077/Measure-partial-liver-volumetric-variations-from-paired-inspiratory-expiratory-chest/10.1117/12.2581077.full}.

\bibitem[Song et~al.(2021)Song, Meng, and Ermon]{song2021denoising}
Jiaming Song, Chenlin Meng, and Stefano Ermon.
\newblock Denoising diffusion implicit models.
\newblock In \emph{International Conference on Learning Representations}, 2021.
\newblock URL \url{https://openreview.net/forum?id=St1giarCHLP}.

\bibitem[Troschel et~al.(2020)Troschel, Troschel, Best, Gaissert, Torriani,
  Muniappan, Seventer, Nipp, Roeland, Temel, and Fintelmann]{Troschel2020}
Amelie~S. Troschel, Fabian~M. Troschel, Till~D. Best, Henning~A. Gaissert,
  Martin Torriani, Ashok Muniappan, Emily E.~Van Seventer, Ryan~D. Nipp,
  Eric~J. Roeland, Jennifer~S. Temel, and Florian~J. Fintelmann.
\newblock Computed tomography-based body composition analysis and its role in
  lung cancer care.
\newblock \emph{Journal of Thoracic Imaging}, 35:\penalty0 91--100, 2020.
\newblock ISSN 15360237.
\newblock \doi{10.1097/RTI.0000000000000428}.

\bibitem[Xu et~al.(2021)Xu, Gao, Khan, Bao, Tang, Deppen, Huo, Sandler,
  Massion, Heinrich, and Landman]{Xu2021}
Kaiwen Xu, Riqiang Gao, Mirza Khan, Shunxing Bao, Yucheng Tang, Steve Deppen,
  Yuankai Huo, Kim Sandler, Pierre Massion, Mattias Heinrich, and Bennett
  Landman.
\newblock Development and characterization of a chest ct atlas.
\newblock volume 11596. SPIE, 2021.
\newblock URL \url{https://doi.org/10.1117/12.2580800}.

\bibitem[Xu et~al.(2022{\natexlab{a}})Xu, Gao, Tang, Deppen, Sandler, Kammer,
  Antic, Maldonado, Huo, Khan, and Landman]{Xu2022}
Kaiwen Xu, Riqiang Gao, Yucheng Tang, Stephen Deppen, Kim Sandler, Michael
  Kammer, Sanja Antic, Fabien Maldonado, Yuankai Huo, Mirza Khan, and
  Bennett~A. Landman.
\newblock Extending the value of routine lung screening ct with quantitative
  body composition assessment.
\newblock page~54. SPIE, 4 2022{\natexlab{a}}.
\newblock ISBN 9781510649392.
\newblock \doi{10.1117/12.2611784}.

\bibitem[Xu et~al.(2022{\natexlab{b}})Xu, Li, Khan, Gao, Antic, Huo, Sandler,
  Maldonado, and Landman]{xu2022body}
Kaiwen Xu, Thomas Li, Mirza~S Khan, Riqiang Gao, Sanja~L Antic, Yuankai Huo,
  Kim~L Sandler, Fabien Maldonado, and Bennett~A Landman.
\newblock Body composition assessment with limited field-of-view computed
  tomography: A semantic image extension perspective.
\newblock 2022{\natexlab{b}}.
\newblock URL \url{https://arxiv.org/abs/2207.06551}.

\end{thebibliography}
\end{document}